\documentclass[a4paper,12pt]{article}

\usepackage{amssymb,amsmath,epsfig}

\newcommand{\A}{\mathcal{A}}
\newcommand{\Ai}{\mathrm{Ai}}
\newcommand{\cl}{\mathrm{cl}}            
\newcommand{\dif}{\mathrm{d}}
\newcommand{\el}{\mathrm{el}}            
\newcommand{\esp}[1]{\mathrm{e}^{#1}}    
\newcommand{\F}{{\cal F}}                
\newcommand{\Hc}{\mathcal{H}}
\newcommand{\kt}{\boldsymbol{k}}
\newcommand{\ls}{\lambda_\mathrm{s}}
\newcommand{\ord}[1]{\mathcal{O}\left(#1\right)}
\renewcommand{\S}{{\cal S}}              
\newcommand{\T}{\mathcal{T}}
\newcommand{\tb}{t_b{}}                  
\newcommand{\U}{\mathcal{U}}
\newcommand{\ui}{\mathrm{i}}             
\newcommand{\xt}{\boldsymbol{x}}

\title{{\bf $\boldsymbol{S}$-matrix and Quantum Tunneling\\ in Gravitational
  Collapse}}

\author{
  M.~Ciafaloni and D.~Colferai\\[1ex]
  \sl Dipartimento di Fisica, Universit\`a di Firenze\\
  \sl and\\
  \sl INFN, Sezione di Firenze, 50019 Sesto Fiorentino, Italy
}

\date{}

\begin{document}

\maketitle

\begin{abstract}
  Using the recently introduced ACV reduced-action approach to transplanckian
  scattering of light particles, we show that the $S$-matrix
  in the region of classical gravitational collapse is related to a tunneling
  amplitude in an effective field space. We understand in this way the role of
  both real and complex field solutions, the choice of the physical ones, the
  absorption of the elastic channel associated to inelastic multigraviton
  production and the occurrence of extra absorption below the critical impact
  parameter. We are also able to compute a class of quantum corrections to the
  original semiclassical $S$-matrix that we argue to be qualitatively sensible
  and which, generally speaking, tend to smooth out the semiclassical results.
\end{abstract}

\vskip 1cm
\begin{minipage}{0.9\textwidth}
\begin{flushright}
  DFF 446/07/08~ \\
\end{flushright}
\end{minipage}
\vskip 1cm

\section{Introduction}

In a recent paper~\cite{ACV07}, Amati, Veneziano and one of us have proposed a
simplified approach to the $S$-matrix for transplanckian scattering of light
particles which could possibly describe the region of classical gravitational
collapse. This treatment originates from the eikonal approach to high-energy
string-gravity proposed in the eighties~\cite{ACV88} and resums a class of
corrections to the leading eikonal operator which correspond to tree diagrams
generated by some effective high-energy graviton emission vertices.

The simplified action approach~\cite{ACV07} is valid in a high energy regime
($Gs\gg\hbar)$) in which the gravitational radius $R\equiv 2G\sqrt s$ and the
impact parameter $b$ of the light scattering particles are large w.r.t. the
string length $\ls$, so that string effects are normally negligible. Thus, the
dynamical variables are basically metric fields which, due to the high-energy
kinematics, are integrated over longitudinal space ($x^{\pm}\equiv x^0\pm x^3$)
and follow a reduced dynamics in transverse space, which is two-dimensional for
$D=4$. The reduced two-dimensional action corresponds to an approximate
shock-wave solution~\cite{ACV93} of the effective gravitational
action~\cite{Li91,Ki95,Ve93}, where however string effects (needed in order to
regularize it and to fully represent the eikonal approach of~\cite{ACV88}) are
neglected.

By solving the classical nonlinear field equations of the reduced action,
ACV~\cite{ACV07} were able to calculate the semiclassical $S$-matrix as function
of $R(s)$ and $b$, down to the strong-coupling regime $b\lesssim R(s)$, where
classical gravitational collapse is expected to occur. Indeed, a basic feature
of the calculation is the existence of a critical value $b_c\sim R$ such that,
for $b<b_c$, the relevant solutions become complex-valued and the eikonal
function shows a power singularity with exponent $3/2$, thus proving the
occurrence of a new regime, possibly related to classical black holes. Indirect
evidence that this is the case comes from an analysis of scattering in more
general axisymmetric configurations\cite{VW08} showing that whenever a
sufficient condition for the existence of closed trapped surfaces (CTS) is
satisfied, the ACV solutions become complex-valued too. Furthermore, the
existence of the critical radius has been confirmed numerically\cite{OM08} by
avoiding the azimuthal averaging approximation of the ACV results, and by
confirming the magnitude of the exponent.

If indeed $b\le b_c$ corresponds to the region of classical gravitational
collapse, the quantum counterpart of it should be related to the fact that, in
this region, the field solutions become complex-valued. By analogy with the
well-known relation of classical to quantum dynamics, this fact suggests that
the $S$-matrix should be related to a quantum tunneling process in a proper
field space. The purpose of the present paper is to investigate this possibility
and to show that indeed a quantum counterpart of the semiclassical calculation
can be found, leading to the tunneling interpretation and to the calculation of
a class of quantum corrections.

The axisymmetric ACV equations are ODE which describe the evolution of some
effective fields in the transverse space variable $r^2$, which plays the role of
time. They involve a scalar field $h(r^2)\equiv 4 d(r^2\dot{\phi})/dr^2$
($\dot{f}\equiv df/dr^2$), which is related to one graviton polarization -- the
other corresponding to soft graviton radiation -- and an auxiliary field
$\rho(r^2)\equiv r^2(1-(2\pi R)^2\dot{\phi})$, in terms of which ACV set the
boundary conditions to their solutions. There are two of them: one is
$\dot{\rho}(\infty)=1$, corresponding to a perturbative behaviour at large
distances, while the other condition is $\rho(0)=0$, which -- because of the
peculiar definition of $\rho$ -- sets to zero a possible $r^2=0$ singularity of
$\dot{\phi}$. ACV argue that the latter condition leads to a consistent
treatment of the $r=0$ boundary, while a singularity of $\dot{\phi}$ would
inficiate the interpretation of the metric coefficient $h_{rr}$.

Given the fact that the semi-classical solutions are found with such boundary
conditions, the two-dimensional dynamics provides a hint to the tunneling
phenomenon. In fact, it provides a Coulomb-like interaction potential
$\sim GsR^2/\rho$ which acts as a barrier, separating the weak field region with
$\rho>0$, $1-\dot{\rho}\sim h< 1$ from the strong field one with
$\rho\le 0, \dot{\phi}>1$. For sufficiently small $b<b_c$, it is impossible --
by real-valued $r^2$-evolution -- to cross such a barrier so as to reach
$\rho(0)=0$, thus avoiding a singular behaviour of the $\dot{\phi}$ field at the
trajectory endpoint. The $S$-matrix instead does it by quantum tunneling,
corresponding to the complex trajectories.

Here, in order to develop the idea just outlined we have to upgrade the
classical level to a quantum level, in which the variable $r^2$ plays the role
of time, and $\rho$ and $\dot{\rho}$ basically satisfy canonical commutation
relations. This sort of quantization is justified in sec.~2, by using a
path-integral interpretation of the reduced action approach, which is recast in
terms of the $\rho$ field only. In sec.~3 we are able to relate the elastic
$S$-matrix (possibly including absorption) to a tunneling amplitude, which is
then expressed at quantum level in terms of a particular choice of Coulomb
wave-functions in $\rho$ space. In sec.~4 we explicitly calculate the ensuing
$S$-matrix at quantum level (including absorption) and we discuss its
semiclassical limits, and the role of quantum corrections. In particular, we
show that elastic unitarity ($|\S_\el|\leq1$) is fulfilled and that the extra
absorption of the collapse-like regime for $b<b_c$ is continuously matched to
normal absorption for $b>b_c$. Finally, in sec.~5, we summarize and discuss our
main results.

\section{The reduced-action approach at semiclassical level\label{s:raa}}

The simplified ACV approach~\cite{ACV07} to transplanckian scattering is based
on two main points.  Firstly, the gravitational field associated to the
high-energy scattering of light particles, reduces to a shock-wave configuration
of the form
\begin{subequations}\label{metrica}
  \begin{align}
    h_{--}\big|_{x^+=0} &= (2\pi R)a(\xt)\delta(x^-) \;, \qquad
    h_{++}\big|_{x^-=0} = (2\pi R)\bar{a}(\xt)\delta(x^+)
    \label{hlong} \\
    h_{ij} &= (\pi R)^2\Theta(x^+ x^-)
    \left(\delta_{ij}-\frac{\partial_i\partial_j}{\nabla^2}\right) h(\xt) \;,
    \label{hij}
  \end{align}
\end{subequations}
where $a$, $\bar{a}$ are longitudinal profile functions, and
$h(\xt)\equiv\nabla^2\phi$ is a scalar field describing one emitted-graviton
polarization (the other, related to soft graviton radiation, is negligible in an
axisymmetic configuration).

Secondly, the high-energy dynamics itself is summarized in the $h$-field
emission-current $\Hc(\xt)$ generated by the external sources coupled to the
longitudinal fields $a$ and $\bar{a}$. Such a vertex has been calculated long
ago~\cite{Li82,ABC89} and takes the form
\begin{equation}\label{vertex}
  -\nabla^2 \Hc \equiv \nabla^2 a \nabla^2\bar{a}-\nabla_i\nabla_j a
  \nabla_i\nabla_j\bar{a} \;,
\end{equation}
which is the basis for the gravitational effective action~\cite{ACV88} from
which the shock-wave solution~\cite{ACV93} emerges. It is directly coupled to
the field $h$ and, indirectly, to the external sources $s$ and $\bar{s}$ in the
reduced 2-dimensional action
\begin{equation}\label{2dimAction}
  \frac{\A}{2\pi Gs} = \int\dif^2 x\left( a\bar{s}+\bar{a}s-\frac12\nabla a
    \nabla\bar{a}+\frac{(\pi R)^2}{2}\left(-(\nabla^2\phi)^2-2\nabla\phi\cdot
\nabla\Hc\right)\right)
\end{equation}
which is the basic ingredient of the ACV simplified treatment.

The equations of motion (EOM) induced by (\ref{2dimAction}) provide, with proper
boundary conditions, some well-defined effective metric fields. The ``on-shell''
action $\A (b,s)$, evaluated on such fields, provides directly the elastic
$S$-matrix
\begin{equation}\label{elSmatrix}
 \S = \exp\left( \frac{\ui}{\hbar} \A(b,s) \right).
\end{equation}
Then, it can be shown~\cite{ACV93,ACV07} that the reduced-action above (where
$R$ plays the role of coupling constant) resums the so-called multi-H diagrams,
contributing a series of corrections $\sim (R^2/b^2)^n$ to the leading eikonal.
Furthermore, the classical field solutions generate an effective metric by
proper estension of~(\ref{metrica}) to the remaining components, as follows
\begin{align}
 \dif s^2 &= -\dif x^+\dif x^
 -\Big[1-\frac{(\pi R)^2}{2} \Theta(x^+ x^-) \nabla^2\phi \Big]
 + 2\pi R \left[ a(z)\delta(x^-)(\dif x^-)^2
 +\bar{a}(z)\delta(x^+)(\dif x^+)^2\right] \nonumber \\
 &\quad -\frac{(\pi R)^2}{4} \nabla^2 \phi
 \left[ |x^+|  \delta(x^-)(\dif x^-)^2+ |x^-| \delta(x^+)(\dif x^+)^2 \right]
 + \dif s_T^2
 \label{dsq} \\
  \dif s_T^2 &= |\dif z|^2+(\pi R)^2\Theta(x^+ x^-)
 \left[ 2|\partial|^2\phi\;|\dif z|^2-\partial^2\phi\;(\dif z)^2
 -{\partial^*}^2\phi\;(\dif z^*)^2 \right]
 \nonumber \\
 &= |\dif z|^2+(\pi R)^2\Theta(x^+ x^-) \left( \delta_{ij} \nabla^2
 - \nabla_i \nabla_j \right)\phi \; \dif x^i \dif x^j \;. \nonumber
\end{align}
where we note that the metric perturbation induced by $h$ has the form of a
gravitational wave with polarization
\begin{equation}\label{TTpol}
 \epsilon^{\mu\nu}_{TT} = (\epsilon_T^{\mu}\epsilon_T^{\nu}
  -\epsilon_L^{\mu}\epsilon_L^{\nu}) \;, \qquad
 \epsilon_L^\mu \equiv (\frac{k^3}{|\kt|},\boldsymbol{0},\frac{k^0}{|\kt|}) \;,
 \qquad \epsilon_T^\mu \equiv (0,\boldsymbol{\epsilon},0) \;.
\end{equation}
The effective metric~(\ref{dsq}) is supposed to be useful in order to bridge the
gap between classical gravitation and the ACV approach. Furthermore, the
$S$-matrix (\ref{elSmatrix}) can be extended to inelastic processes on the basis
of the same emitted-graviton field $h(\xt)$.

In the case of axisymmetric solutions, where $a=a(r^2)$, $\bar{a}=\bar{a}(r^2)$,
$\phi=\phi(r^2)$ it is straightforward to see, by using eq.~(\ref{vertex}), that
$\dot{\Hc}(r^2)=\dot{a}\dot{\bar{a}}$ becomes proportional to the $a,\bar{a}$
kinetic term. Therefore, the action~(\ref{2dimAction}) can be rewritten in the
more compact one-dimensional form
\begin{equation}\label{1dimAction}
  \frac{\A}{2\pi Gs}=\int\dif r^2\left( a(r^2)\bar{s}(r^2) + \bar{a}(r^2)s(r^2)
    -2\rho\dot{\bar{a}}\dot{a} - \frac{2}{(2\pi R)^2}(1-\dot\rho)^2\right) \;,
\end{equation}
where we have introduced the auxiliary field $\rho(r^2)$
\begin{equation}\label{rho}
  \rho=r^2\big(1-(2\pi R)^2\dot\phi\big) \;, \qquad
  h=4\dot{(r^2\dot\phi)}=\frac1{(\pi R)^2}(1-\dot\rho)
\end{equation}
which incorporates the $\phi$-dependent interaction. The external sources
$s(r^2)$, $\bar{s}(r^2)$ are assumed to be axisymmetric also, and are able to
describe the particle-particle case by setting $\pi s(r^2)=\delta(r^2)$,
$\pi\bar{s}(r^2)=\delta(r^2-b^2)$, where the azimuthal averaging procedure of
ACV is assumed.%
\footnote{The most direct interpretation of this configuration is the scattering
  of a particle off a ring-shaped null matter distribution, which is
  approximately equivalent to the particle-particle case by azimuthal
  averaging~\cite{ACV07}.}

The equations of motion, specialized to the case of particles at impact
parameter $b$ have the form
\begin{align}
  \dot{a} &= -\frac1{2\pi\rho} \;, \qquad
  \dot{\bar{a}} = -\frac1{2\pi\rho}\Theta(r^2-b^2) \;, \label{eoma}\\
  \ddot{\rho} &= \frac1{2\rho^2}\Theta(r^2-b^2) \;, \qquad
  \dot{\rho}^2+\frac1{\rho} = 1 \qquad (r > b) \label{eomrho}
\end{align}
and show the repulsive ``Coulomb'' potential in $\rho$-space (mentioned in the
introduction), which acts for $r>b$ and will play an important role in the
following. By replacing the EOM~(\ref{eoma}) into eq.~(\ref{1dimAction}), the
reduced action can be expressed in terms of the $\rho$ field only, and takes the
simple form
\begin{equation}\label{rhoAction}
  \frac{\A}{Gs} = -\int\dif r^2
  \left(\frac1{R^2}(1-\dot\rho)^2-\frac1{\rho}\Theta(r^2-b^2)\right) \;,
\end{equation}
which is the one we shall consider at quantum level.

Let us now recall the main features of the classical ACV solutions of
eq.~(\ref{eomrho}). First, we set the ACV boundary conditions
$\dot\rho(\infty)=1$ (matching with the perturbative behaviour), and
$\rho(0)=0$, where the latter is required by a proper treatment~\cite{ACV07} of
the $r^2=0$ boundary.%
\footnote{A nonvanishing $\rho(0)$ would correspond to some outgoing flux of
  $\nabla \phi$ and thus to a $\delta$-function singularity at the origin of
  $h$, which is not required by external sources.}
Then, we find the Coulomb-like solution
\begin{align}
  \rho &= R^2\cosh^2\chi(r^2) \;, \qquad \dot\rho = \sqrt{1-\frac1{\rho}}
   = \tanh\chi(r^2) \qquad (r^2 \geq b^2) \nonumber \\
  r^2 &= b^2 + R^2(\chi+\sinh\chi\cosh\chi-\chi_b-\sinh\chi_b\cosh\chi_b) \;,
  \label{clSol}
\end{align}
to be joined with the behaviour $\rho=\dot\rho(b^2)r^2$ for $r^2\leq b^2$. The
continuity of $\rho$ and $\dot{\rho}$ at $r^2=b^2$ requires the matching
condition
\begin{equation}\label{crit}
 \rho_b = b^2 \tanh\chi_b = R^2\cosh^2\chi_b \;, \qquad
 \frac{R^2}{b^2} = t_b(1-t_b^2) \;,
\end{equation}
which acquires the meaning of criticality equation.

Indeed, if the impact parameter $b^2$ exceeds a critical value
$b_c^2=(3\sqrt{3}/2)R^2$ at which eq.~(\ref{crit}) is stationary, real valued
solutions of type~(\ref{clSol}) with the above boundary condition do exist,
while for $b<b_c$ they become complex-valued. For $b<b_c$, the class of
real-valued solutions of type~(\ref{clSol}) has
$\rho(0)=\rho(b^2)-b^2\dot\rho(b^2)>0$, and for a particular initial value
$\chi(0)=\chi_m$ we reach the minimal $\rho_m$, determined by
\begin{equation}\label{chimin}
  \frac{b^2}{2R^2}=\cosh^3\chi_m \sinh\chi_m=\frac{t_m}{(1-t_m^2)^2} \;.
\end{equation}
Furthermore, the action~(\ref{rhoAction}) evaluated on the equation of motion
becomes
\begin{equation}\label{eomAction}
 \frac{\A}{Gs} = \log(4L^2) - \log\frac{1+t_b}{1-t_b} + 1
 -\frac{b^2}{R^2}(1-t_b^2) \;, \qquad (t_b \equiv \tanh\chi_b)
\end{equation}
and provides directly the $b$-dependent eikonal occurring in the elastic
$S$-matrix.

\begin{figure}[!ht]
  \centering
  \includegraphics[width=0.45\textwidth]{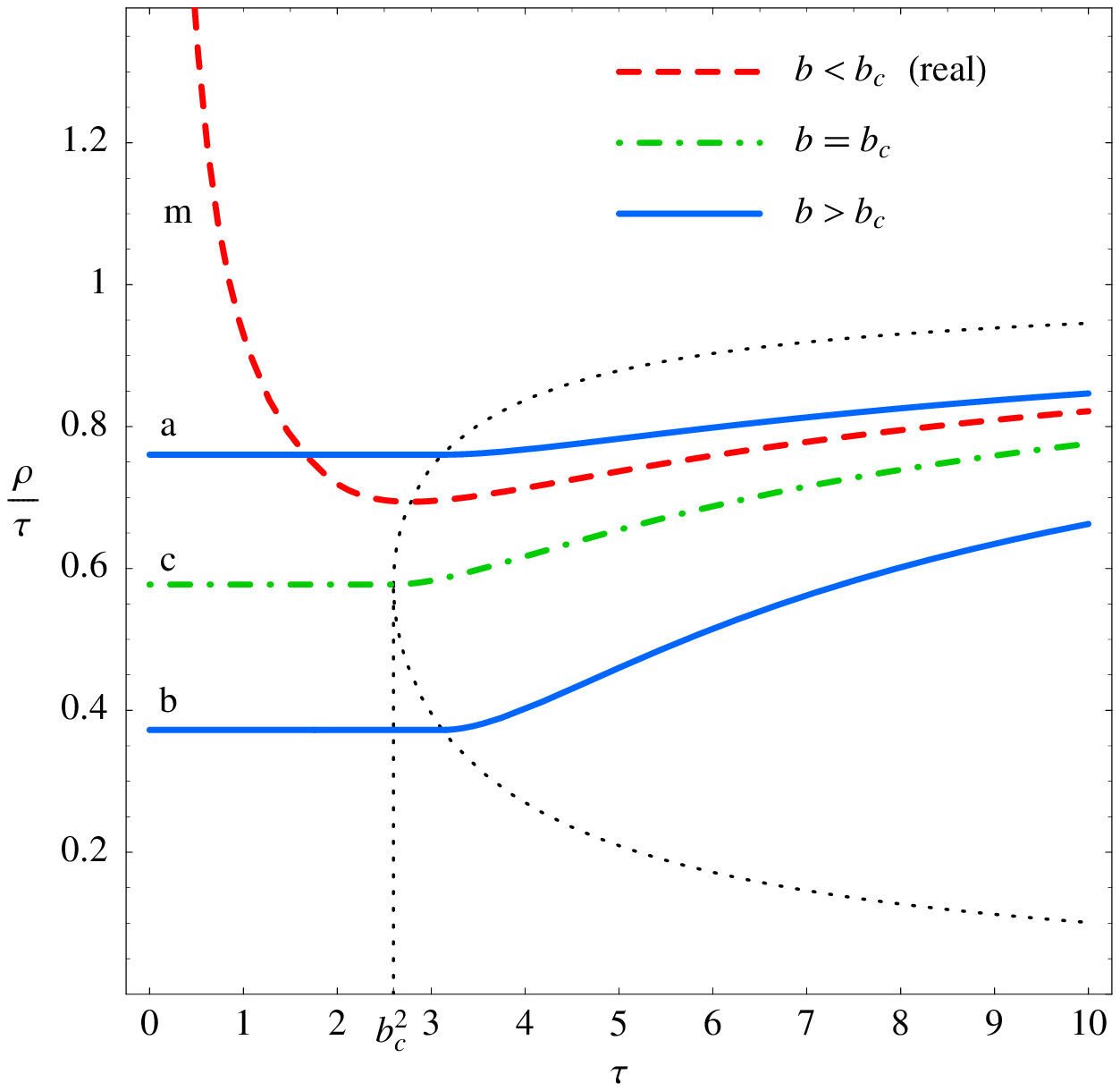}\hfill
  \includegraphics[width=0.45\textwidth]{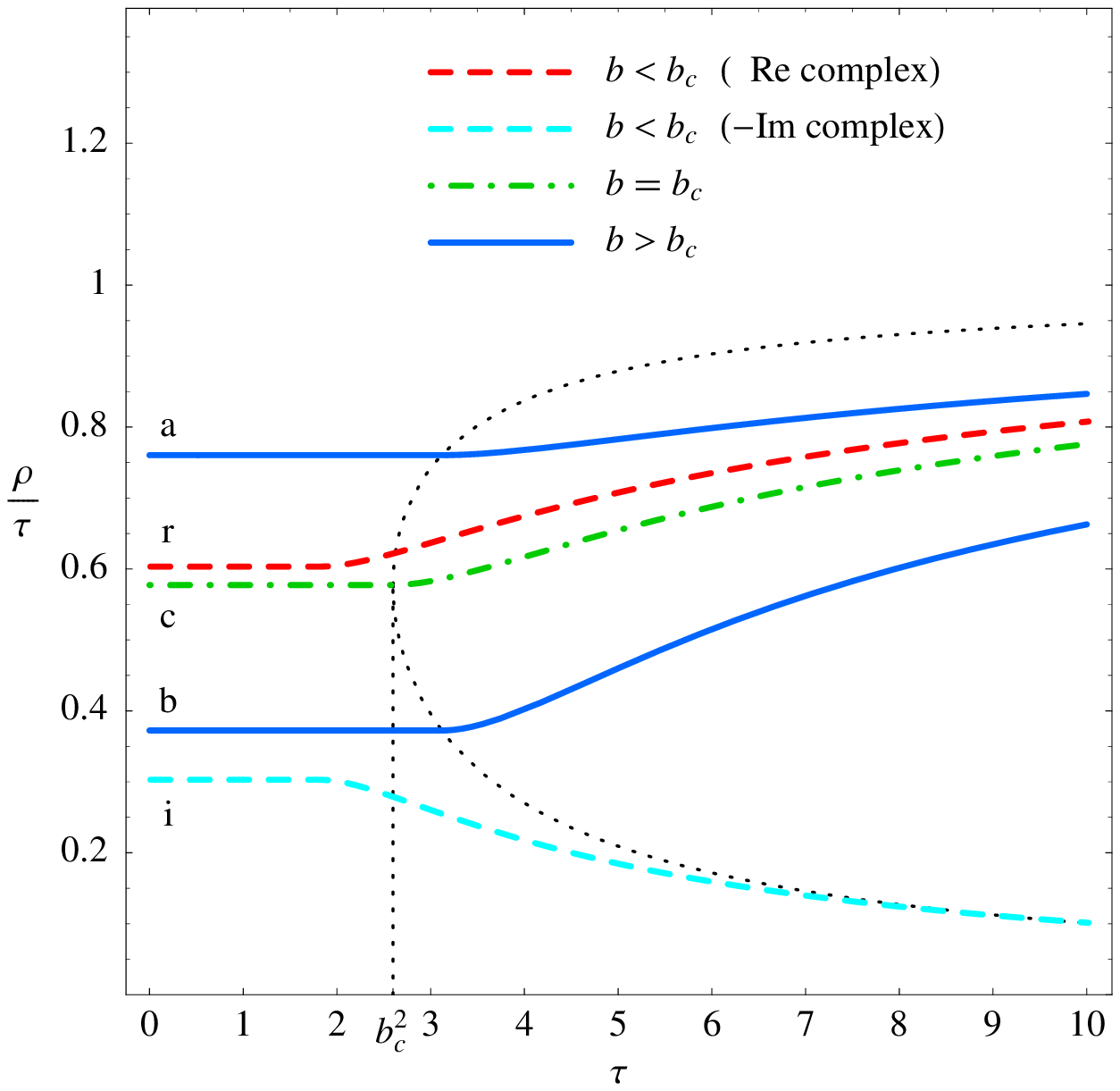}
  \caption{\it Semiclassical solutions $\rho(\tau)/\tau$ showing: (a,b)
    supercritical branches (solid-blue) at $b^2=1.2\,b_c^2$, (c) the critical
    one (dash-dotted green) and subcritical ones at $b^2=0.7\,b_c^2$. In the
    latter case we contrast (m) the singular real-valued one (dashed-red on the
    left) with (r,i) the regular complex-valued one (dashed, on the right). The
    black-dotted contour shows the border between free and Coulomb-like
    evolution for $b\geq b_c$.}
  \label{f:classicalSol}
\end{figure}

The various branches of physical solutions for
$\rho(r^2)/r^2= 1-(2\pi R)^2\dot{\phi}$ are pictured in
fig.~\ref{f:classicalSol}.  We see that, for $b\geq b_c$, there are two
solutions with everywhere regular $\phi$ field, one of them matching the
iterative solution. On the other hand, for $b<b_c$ the regular solutions become
complex valued (fig.~\ref{f:classicalSol}b). They are compared in
fig.~\ref{f:classicalSol}a with the irregular real-valued ones which have
$\rho(0)=R^2\cosh^2(\chi_m)(1-2\sinh^2\chi_m)>0$.
Due to the definition of $\rho=r^2[1-(2\pi R)^2\dot\phi]$, which has the
kinematical factor $r^2$, we see that such solutions show a singularity of the
$\dot\phi$ field of type $\dot\phi\simeq-\rho(0)/r^2<0$, so that the metric
coefficient $h_{rr}$ must change sign at some value of $r^2\sim R^2$ and is
singular at $r=0$.

A clearcut interpretation of the (unphysical) real-valued solutions with $b<b_c$
and $\rho(0)>0$ is not really available yet. However, we know that in about the
same impact parameter region classical CTS do exist, as shown
in~\cite{EG02,KV02,VW08}. It is therefore tempting to guess that such field
configurations of the ACV approach (which are singular and should have
negligible quantum weight) correspond to classically trapped surfaces. In this
picture, the complex-valued solutions with $\rho(0)=0$ (which are regular, and
should have finite quantum weight) would correspond to the tunneling transition
from the perturbative fields with $\dot{\rho}(\infty)=1$ and positive $\rho$ to
the ``untrapped'' configuration with $\rho(0)=0$.

In order to develop the above suggestion, in the following we shall consider the
action~(\ref{rhoAction}) at quantum level, by defining the $S$-matrix as the
path-integral over $\rho$-field configurations induced by that action.

\section{The quantum level and tunneling amplitude}

The idea is to introduce the quantum $S$-matrix as a path-integral in
$\rho$-space of the reduced-action exponential. In this ``sum over actions''
interpretation the semiclassical limit will automatically agree with the
expression in eq.~(\ref{rhoAction}) above, which is based on the ``on-shell''
action. Furthermore, calculable quantum corrections will be introduced.

\subsection{Quantized elastic $\boldsymbol{S}$-matrix}

Following the above suggestion, and neglecting absorptive effects induced by the
$h$-field, we define
\begin{equation}\label{Spi}
  \S_\el(b,s) = \int_{
    \begin{matrix} _{\rho(0)=0} \\ ^{\dot\rho(\infty)=1\;\,} \end{matrix}
  }
  [\mathcal{D}\rho(\tau)] \;
 \exp\left\{-\frac{\ui}{\hbar}\int\dif\tau \; L(\rho,\dot\rho,\tau)\right\}
\end{equation}
where we use the expression~(\ref{rhoAction}) of the reduced action, with the
notation $\tau\equiv r^2$ and we introduce the Lagrangian
\begin{equation}\label{lagrangian}
 L(\rho,\dot\rho,\tau) = \frac1{4G}\left[(1-\dot\rho)^2
 - \frac{R^2}{\rho}\Theta(\tau-b^2)\right] \;,
\end{equation}
with the boundary conditions $\rho(0)=0,\;\dot\rho(\infty)=1$ introduced by ACV
and discussed in sec.~\ref{s:raa}.

The definition~(\ref{Spi}) given above is equivalent, by a Legendre transform
and use of the Trotter formula~\cite{RS72}, to quantize the $\tau$-evolution
Hamiltonian $H(\tau)$ to be introduced shortly, and to calculate the evolution
operator $\U(0,\infty)$, thus reducing the $S$-matrix calculation to a known
quantum quantum-mechanical problem.  In fact, by eq.~(\ref{lagrangian}), we can
introduce the ``conjugate momentum''
\begin{equation}\label{conjMom}
  \Pi \equiv \frac{\partial L}{\partial\dot\rho} = \frac1{2G}(\dot\rho-1)
\end{equation}
and we obtain
\begin{equation}\label{hamiltonian}
  H(\tau) \equiv \Pi\dot\rho - L = \frac1{4G}\left((\dot\rho)^2-1
  +\frac{R^2}{\rho}\Theta(\tau-b^2)\right) \;, \qquad \dot\rho = 1+2G{\Pi}
\end{equation}
from which the classical EOM~(\ref{eomrho}) can be derived. Then, quantizing the
evolution according to eq.~(\ref{Spi}) amounts to assume the canonical
commutation relation
\begin{equation}\label{cr}
  [\rho,\Pi] = \ui\hbar \;, \qquad
  \dot\rho = -2{\ui\hbar}G\frac{\partial}{\partial\rho}
  \equiv -\frac{\ui R^2}{2\alpha}\frac{\partial}{\partial\rho} \;, \qquad
  \alpha\equiv\frac{Gs}{\hbar}
\end{equation}
and to quantize the Hamiltonian~(\ref{hamiltonian}) accordingly:
\begin{equation}\label{Hquant}
 \frac{\hat{H}}{\hbar} = -\frac{R^2}{4\alpha} \frac{\partial^2}{\partial\rho^2}
 + \alpha\left(\frac{\Theta(\tau-b^2)}{\rho}-\frac1{R^2}\right)
 \equiv \frac{H_0}{\hbar}+\frac{\alpha}{\rho}\Theta(\tau-b^2) \;.
\end{equation}
Finally, the path-integral~(\ref{Spi}) for the $S$-matrix without absorption is
related by Trotter's formula to a tunneling amplitude involving the
time-evolution operator $\U(0,\infty)$:
\begin{equation}\label{tunnel}
 \S(b,s)\sim \T(b,\alpha) \equiv \langle\rho=0|\U(0,\infty)|\Pi=0\rangle
 \;, \qquad H_0 |\Pi=0\rangle = 0 \;
\end{equation}
where $\U(\tau,\infty)$ is calculated with $\tau$-antiordering and the
normalization of states will be fixed below.

We note that the commutation relation~(\ref{cr}) does not follow from first
principles, but is simply induced by the path-integral definition (\ref{Spi}).
Note also that here we allow fluctuations in transverse space, but we keep
frozen the shock-wave dependence on the longitudinal variables $x^{\pm}$. This
means that our account of quantum fluctuations is admittedly incomplete and
should be considered only as a first step towards the full quantum level. This
step, defined by~(\ref{Spi})-(\ref{tunnel}), has nevertheless the virtue of
reproducing the semiclassical result for $\alpha\to\infty$. Furthermore, we
think that such commutation relations are of the expected order of magnitude. An
argument in this direction is to notice that the shock-wave of ACV has
wavefronts%
\footnote{ The second wavefront is better seen by replacing $h(\xt)$ in
  eq.~(\ref{hij}) by $\int\frac{\dif^2\kt}{(2\pi)^2} \;\tilde{h}(\kt)
  \esp{\ui\kt\cdot\xt}J_0(|\kt|\sqrt{x^+ x^-})$, as argued in~\cite{ACV07}.}
at $x^+x^-=0$ and $x^+x^- \simeq r^2$, and the latter implies
$t\simeq-(|x_3|+r^2/\alpha)$ at large negative times of order $\alpha$.
Therefore, time- and $r^2$-derivatives turn out to be related by a factor of
$1/\alpha$ as a consequence of a sort of retardation in the occurrence of that
``precursor'' wavefront, a factor which also occurs in the commutation relation
(\ref{cr}). Finally, corrections of relative order
$(\hbar/Gs)R^2/b^2=\lambda_P^2/b^2$ are expected here, and at full quantum level
also.

\subsection{Inelastic $\boldsymbol{S}$-matrix and absorption}

We have so far defined the $S$-matrix as if no inelastic processes were present.
However, the very existence of the emitted graviton field $h$ implies the
existence of multi-graviton production and, by unitarity, of absorption in the
elastic channel. Therefore, in order to find a unitary $S$-matrix at quantum
level, it is essential to introduce both phenomena. At semiclassical level, ACV
noticed that the classical $h$-field solution
\begin{equation}\label{hcl}
  h_\cl(\tau) = \frac1{(\pi R)^2} \big(1-\dot\rho_\cl(\tau)\big)
\end{equation}
induces in the eikonal formulation an inelastic $S$-matrix which is
approximately described by the coherent state operator
\begin{align}
 &\S = \exp\left( \frac{\ui}{\hbar} \A(b,s) \right)
 \exp\left( \frac{2\ui\sqrt{\alpha}}{\pi R}
 \int_0^\infty\dif^2 x\; \big(1-\dot\rho_\cl(\tau)\big) \Omega(\xt) \right)
 \label{Scs} \\ \nonumber
 &\Omega(\xt) \equiv \int\frac{\dif^2\kt\,\dif k_3}{2\pi\sqrt{k_0}}
 \left[ a(\kt,k_3) \esp{\ui\kt\cdot\xt} + h. c. \right]
 \equiv A(\xt)+A^\dagger(\xt))\;,\\
 &[ A(\xt), A^\dagger(\xt')] = Y_b \delta(\xt - \xt')
 \label{Omega}
\end{align}
where the operator $\Omega(\xt)$ incorporates both emission and absorption of
the $h$-fields and $Y_b$ parametrizes the ($b$-dependent) rapidity phase space
which is effectively%
\footnote{When energy conservation is taken into account~\cite{CV08}, absorptive
  corrections consistent with the AGK cutting rules~\cite{AGK73} suppress the
  fragmentation region in a $b$-dependent way, so as to yield a purely central
  process for $b\sim b_c$.}
allowed by energy conservation~\cite{CV08} . By normal ordering of~(\ref{Scs}),
we derive the semiclassical elastic-channel absorption factor
\begin{equation}\label{absorption}
  {\cal J}\equiv |\S_\el(b,s)|= \esp{-\frac{2Y_b \alpha}{\pi}\int\dif\tau\;
 (1-\dot\rho_\cl)^2} \qquad
\end{equation}
which is dependent, of course, on the classical solution singled out at
semiclassical level.

A possible way to extend the coherent state definition~(\ref{Scs}) to the
quantum level, is just to introduce it in the path-integral
formulation~(\ref{Spi}) as follows
\begin{equation}\label{inSpi}
  \S(b^2,s; \Omega] = \int_{
    \begin{matrix} _{\rho(0)=0} \\ ^{\dot\rho(\infty)=1\;\,} \end{matrix}
  }
  [\mathcal{D}\rho(\tau)] \;
 \esp{-\ui\int\dif\tau \; L(\rho,\dot\rho,\tau)} \;
 \esp{\frac{2\ui\sqrt{\alpha}}{\pi R}\int\dif^2\xt\;
 [1-\dot\rho(\tau)]\Omega(\xt)} \;,
\end{equation}
where $\Omega(\xt)$ acts on the multi-graviton Fock space, but is to be
considered as a c-number current with respect to the quantum variables
$\rho,\dot{\rho}$.

In the elastic channel, the $\Omega$-dependent exponential in~(\ref{inSpi}) is
to be replaced by its vacuum expectation value
$\exp[-\frac{2Y_b \alpha}{\pi}\int\dif\tau\; (1-\dot\rho)^2]$.  Of course, in
this quantum extension of (\ref{absorption}), no commitment is made to a
particular classical solution so that the output will presumably contain a
weighted superposition of the various classical paths satisfying the boundary
conditions, that we shall calculate in the following.

\subsection{Tunneling amplitude neglecting absorption}

With the above warnings about the meaning of quantization and the role of
inelasticity, let us derive a more detailed expression of the tunneling
amplitude~(\ref{tunnel}) without absorption
\begin{equation}\label{tunnel_2}
 \T(b,\alpha) \equiv \langle\rho=0|\U(0,\infty)|\Pi=0\rangle
 =\langle\rho=0|\psi(\tau=0)\rangle \;, \qquad H_0 |\Pi=0\rangle = 0 \;.
\end{equation}
where the initial (final) state expresses the boundary condition
$\dot\rho(\infty)=1$ ($\rho(0)=0$), $\psi(\tau)$ is the time-dependent wave
function, and $\U(\tau,\infty)$ is the evolution operator in the Schr\"odinger
picture, calculated with $\tau$-antiordering (according to the suggestion above
that true time is related to $-r^2/\alpha$).

Since the Hamiltonian~(\ref{hamiltonian}) is time-dependent, the expression of
the wave function at time $\tau\equiv r^2$ is related to the evolution due to
the Coulomb Hamiltonian $H_c\equiv H_0+Gs/\rho$ by
\begin{align}
 |\psi(\tau)\rangle &= 
 \exp\left(\frac{-\ui H_c \tau}{\hbar}\right) \U_c(0,\infty) |\Pi=0\rangle
 \qquad (\tau\geq b^2) \label{psiIg} \\
 &=  \exp\left(\frac{\ui H_0 (b^2-\tau)}{\hbar}\right)
 \exp\left(\frac{-\ui H_c b^2}{\hbar}\right) \U_c(0,\infty) |\Pi=0\rangle
 \qquad (\tau < b^2) \label{psiIl} \;.
\end{align}
where, according to eq.~(\ref{Hquant}), we have used ``free'' evolution for
$\tau<b^2$.  Therefore, the tunneling amplitude can be related to a Coulomb wave
function as follows
\begin{align}
 \T(b,\alpha) &= \langle\rho=0|\psi(0)\rangle = \langle\rho=0|
 \exp\left(\frac{\ui H_0 b^2}{\hbar}\right)
 \exp\left(\frac{-\ui H_c b^2}{\hbar}\right) \U_c(0,\infty) |\Pi=0\rangle
 \nonumber \\
 &= \int\frac{\dif\rho}{(\pi b^2/\ui\alpha)^{1/2}}\;
 \esp{-\ui\alpha(\rho^2/b^2+b^2)}\psi_c(\rho)
 \label{Tval}
\end{align}
where we note the free Gaussian propagator, acting on the continuum Coulomb wave
function
\begin{equation}\label{psic}
  \psi_c(\rho) \equiv \langle\rho|\U_c(0,\infty)|\Pi=0\rangle \;.
\end{equation}
The latter, due to the infinite evolution from the initial condition
$\Pi=0 \iff \dot\rho=1$, is a solution of the stationary Coulomb problem
\begin{equation}\label{stCoul}
 H_c \psi_c(\rho) = \hbar\left[-\frac{1}{4\alpha}\frac{\dif^2}{\dif\rho^2}
 +\alpha\Big(\frac1\rho-1\Big)\right]\psi_c(\rho) = 0 \;.
\end{equation}
with zero energy eigenvalue (where from now on we express $\rho, r^2, b^2$ in
units of $R^2=4G^2s$).  The form of $\psi_c(\rho)$ is better specified by the
Lippman-Schwinger equation
\begin{equation}\label{LSeq}
 \psi_c(\rho) = \esp{2\ui\alpha\rho} + \alpha G_0(0) \frac1{\rho} \psi_c(\rho)
 \;, \qquad G_0(E) = [E-H_0+\ui\epsilon]^{-1}
\end{equation}
and thus contains an incident wave with $\dot\rho=1$, plus a reflected wave for
$\rho>0$ and a transmitted wave in the $\rho<0$ region.

We then conclude that the amplitude~(\ref{tunnel}) is, by eq.~(\ref{Tval}), the
convolution of a gaussian propagator with the Coulomb wave function
$\psi_c(\rho)$, which has a tunneling interpretation with the Coulomb barrier.
In fact, by eq.~(\ref{LSeq}), it contains a transmitted wave in $\rho<0$ (where
the Coulomb potential is attractive) and incident plus reflected waves in
$\rho>0$ (where it is repulsive).

Note that, at $b=0$ we simply have $\T(0,\alpha)=\psi_c(0)$, so that the
tunneling interpretation is direct. On the other hand for $b>0$, the convolution
with the free propagator changes the problem considerably, and is the source of
the critical impact parameter, as we shall see below.

\section{Calculation of quantum tunneling amplitude}

We shall now proceed to the actual calculation of the tunneling amplitude
(\ref{Tval}) without absorption in terms of the wave function (\ref{psic}). We
shall then introduce absorption according to the definition (\ref{inSpi}), by
discussing in particular the $S$-matrix in the elastic channel.

\subsection{The tunneling wave function and $\boldsymbol{b=0}$ case}

The explicit solution of (\ref{stCoul}) is given by a particular confluent
hypergeometric function of $z\equiv -4\ui\alpha\rho$ defined as follows
\begin{align}
  \psi_c &= N_c \, z \, \esp{-z/2} \Phi(1+\ui\alpha, 2,z) \;, \qquad
  z\Phi''+(2-z)\Phi'-(1+\ui\alpha)\Phi=0
  \nonumber \\
  \Phi&\simeq z^{-(1+\ui\alpha)}\big(1+O(1/z)\big) \;,
 \qquad(\ui z\sim\rho\rightarrow-\infty)
 \label{hyper}
\end{align}
where $\Phi$ is defined in terms of its asymptotic power behaviour for
$\rho\rightarrow -\infty$ and the normalization factor $N_c$, to be found below,
is chosen so as to have, asymptotically, a pure-phase incoming wave for
$\rho\simeq L^2\gg 1$, $L^2$ being an IR parameter used to factorize the Coulomb
phase. We shall call this prescription as the ``Coulomb phase'' normalization at
$b=\infty$.

Here we note that the value $c=2$ in $\Phi(1+\ui\alpha, c,z)$ yields a
degenerate case for the differential equation in (\ref{hyper}) in which the
standard solution with the $\rho\to-\infty$ outgoing wave, usually called
$U(1+\ui\alpha, 2, z)$~\cite{AS}, develops a $z=0$ singularity of the form
$A/z+B\log z$. Then, the continuation to $\rho>0$ is determined by requiring the
continuity of wave function and its {\it flux} at $\rho=0$, as is appropriate
for a principal part determination of the ``Coulomb'' singularity. The outcome
involves therefore an important contribution at $\rho>0$ of the regular solution
$F(1+\ui\alpha, 2, z)$, so that we obtain
\begin{align}\label{wavef}
 z\esp{-z/2}\Phi &= z \esp{-z/2} \left(U(1+\ui\alpha,2,z)
 +\frac{\ui\pi\Theta(\ui z)}{\Gamma(\ui\alpha)}F(1+\ui\alpha,2,z)\right)
 \\ \nonumber
 &\simeq \esp{(\pi\alpha-z/2)}\cosh(\pi\alpha)z^{-\ui\alpha}
 +\frac{\Gamma(-\ui\alpha)}{\Gamma(\ui\alpha)}\esp{(\pi\alpha+z/2)}
 \sinh(\pi\alpha) (-z)^{\ui\alpha} \qquad (\ui z\to +\infty)
\end{align}
We are finally able to determine the normalization factor $N_c$ and the value of
$\psi_c(0)$, which is finite and non-vanishing, as follows
\begin{equation}\label{psi_0}
 \T(0,\alpha) = \psi_c(0) = \frac{N_c}{\Gamma(1+\ui\alpha)}
 =(4\alpha L^2)^{\ui\alpha}
 \frac{\exp(-\pi\alpha/2)}{\Gamma(1+\ui\alpha)\cosh\pi\alpha}
\end{equation}
a value which is of order $\esp{-\pi\alpha}$, the same order as the wave
transmitted by the barrier.

In the $b=0$ case, the tunneling amplitude is simply the value $\psi_c(0)$ in
(\ref{psi_0}), with the normalization just discussed, and is of order
$\exp(-\pi\alpha)$, as our $S$-matrix, leaving aside the coherent state
describing multigraviton production.  We obtain, from ACV and from
eq.~(\ref{psi_0}), a more detailed expression in terms of the known on-shell
action (\ref{eomAction}), as follows
\begin{align}
 \A (0,s) &= \alpha\left(\log \frac{4L^2}{R^2}+1+\ui\pi\right) \;,\\ \nonumber
 \T(0,\alpha) &= \exp\big\{\ui\A (0,s)\big\}\big(1+\ord{1/\alpha}\big) \;.
\end{align}
We see that the tunneling amplitude provides the semiclassical result for the
elastic $S$-matrix and quantum corrections to it. It also determines, in a way
to be discussed shortly, both normal graviton radiation and the corresponding
absorption.

\subsection{Integral representation of tunneling amplitude at
  $\boldsymbol{b>0}$}
 
For $b>0$, the calculation of $\T$ in (\ref{Tval}) involves a nontrivial
integral, which should be investigated with care. A preliminary analysis can be
performed in the WKB approximation, which is straightforward. In fact, by
setting $\rho=\cosh^2\chi$, we have
\begin{equation}\label{WKB}
 \psi_{\mathrm{WKB}}^c\sim \exp\left(\frac{\ui}{\hbar}\int_1^{\rho} \dif\rho' \;
 \sqrt{1-\frac{1}{\rho'}}\right)= \esp{2\ui\alpha(\sinh \chi\cosh\chi-\chi)}
\end{equation}
and the integral in (\ref{tunnel}) is dominated by a stationarity point at
\begin{equation}\label{critical}
 \sqrt{1-1/\rho}-\rho/b^2=0=R^2/b^2-t_b(1-t_b^2)\qquad
 (\text{criticality equation})
\end{equation}
yielding the phase
\begin{equation}\label{WKBphase}
 \frac{\ui}{2\hbar}(2\sinh\chi_b\cosh\chi_b-2\chi_b
 -\frac{\cosh^4\chi_b}{b^2}-b^2)
 = \ui Gs(-2\chi_b-\frac{\cosh\chi_b}{\sinh\chi_b})
\end{equation}
which reproduce both the criticality equation and the on-shell action of ACV,
apart from an overall phase, which has not been determined in the expression
(\ref{WKB}).

The detailed calculation of $\T$, yielding quantum corrections also, is
done by using standard Fourier-type representations for $U$ and $F$ \cite{AS} in
eq.~(\ref{wavef}) and by performing the gaussian $\rho$-integral in
eq.~(\ref{Tval}). We then obtain the integral representation
\begin{equation}\label{int-rep}
 \frac{\T(b,\alpha)}{\T(0,\alpha)} = 2 b^2 \ui\alpha \left[-\int_1^{\infty}
 \dif t \; t \left(\frac{t-1}{t+1}\right)^{\ui\alpha}\esp{\ui\alpha b^2(t^2-1)}+
 \int_C \dif t\;\frac{t}{2}\left(\frac{t-1}{t+1}\right)^{\ui\alpha}
 \esp{\ui\alpha b^2(t^2-1)}\right]
\end{equation}
where $C$ is a contour encircling the branch points at $t=\pm 1$ in the
anticlockwise direction. In eq.~(\ref{int-rep}), the $b=0$ limit is accounted
for by the first term, while the second takes over for finite $b$ values, and is
leading by a factor of $\sinh(\pi\alpha)$ for $b\gg 1$. The latter factor is
exhibited by rewriting the contour integral in terms of the discontinuity of the
integrand for $-1<t<1$.

More precisely, for sizeable values of $b^2$, such that
$\exp(\alpha b^{2/3})\gg 1$, eq.~(\ref{int-rep}) is dominated by the second term
which, by closing the contour on the branch-cut, takes the form
\begin{align}
 \T(b,\alpha) &\simeq
 \frac{(\ui\alpha/\esp{})^{\ui\alpha}}{\Gamma(\ui\alpha)}
 \tanh(\pi\alpha) \,2 b^2 \int_{-1}^{+1} \dif t\; t\,
 \esp{\ui\alpha\F(t,b^2)} \;, \qquad (\esp{\alpha b^{2/3}} \gg 1 )
 \label{Trep} \\
 \F(t,b^2) &\equiv \log(4L^2)+1-\log\frac{1+t}{1-t}-b^2(1-t^2)
 \label{phase}
\end{align}
where, as already noticed, the phase $\F$ is closely related to the original
action (\ref{eomAction}) when evaluated at the values $t=t_b$ satisfying the
criticality condition.

It is interesting to note that the amplitude expression (\ref{int-rep})
resembles the ``sum over solutions'' interpretation of the $S$-matrix in ACV, by
setting $t=t_b\equiv\tanh\chi_b$ and by choosing a proper measure factor. In
fact, with this identification, the integrand carries the same phase as
eq.~(\ref{WKBphase}) given before. Keep in mind, however, that the full
semiclassical expression of the ACV $S$-matrix carries also a coherent state
factor yielding graviton radiation and the corresponding absorption of the
elastic amplitude, given in terms of the field $h(r^2)=(1-\dot{\rho}(r^2)/(\pi
R)^2$, for each value of $t_b$. We think, therefore, that the expression
(\ref{int-rep}) --- valid as it stands for the ``elastic'' part of the action
(\ref{rhoAction}) --- should be improved in order to extend it to inelastic
processes and to take into account this effect.

\subsection{Including absorption at quantum level\label{s:iaql}}

In order to take into account multi-graviton emission, the $S$-matrix should be
defined as in eq.~(\ref{inSpi}). By limiting ourselves to the elastic channel,
we should calculate the normal-ordering suppression factor analogous
to~(\ref{absorption}), and we obtain
\begin{align}\label{Sel}
 \S_\el(b^2,s) &= \langle0|\S|0\rangle = N
 \int_{\begin{matrix} _{\rho(0)=0} \\ ^{\dot\rho(\infty)=1\;\,} \end{matrix}}
  [\mathcal{D}\rho(\tau)] \;
 \esp{-\frac{\ui}{\hbar}\int\dif\tau \; L(\rho,\dot\rho,\tau)} \;
 \esp{-\alpha y \int\dif\tau \; \left(1-\dot\rho(\tau)\right)^2} \;,
\end{align}
where the parameter $y \equiv 2Y_b/\pi$ effectively takes into account the
longitudinal phase space, limited by energy conservation~\cite{CV08}.

The absorption term in~(\ref{Sel}) adds an imaginary part to the kinetic term in
the Lagrangian and formally changes the definition of the Hamiltonian and of the
quantization condition in terms of a parameter
$\tilde\alpha\equiv\alpha(1-\ui y)$
\begin{equation}\label{Htilde}
 \tilde{H} = \tilde\alpha\left(\dot{\hat{\rho}}\,^2-1\right)+\frac{\alpha}{\hat\rho}
 \Theta(\tau-b^2) \;, \qquad
 [\hat\rho,\dot{\hat{\rho}}]=\frac{\ui\hbar}{2\tilde\alpha} \;, \qquad
 \tilde\alpha \equiv \alpha(1-\ui y) \;.
\end{equation}
A simple way to take into account such changes is to solve the evolution
equation for the wave-function
$\langle t|\tilde\psi(\tau)\rangle\equiv\psi(t;\tau)$ directly in the
momentum representation in which $\dot{\hat{\rho}}=t$ is diagonal. We simply
obtain
\begin{equation}\label{momSpaceEq}
 \ui\frac{\partial}{\partial\tau}\psi(t;\tau) = \left[\tilde\alpha(t^2-1)
 +\alpha\Theta(\tau-b^2)\left(\frac{\ui}{2\tilde\alpha}
 \frac{\partial}{\partial t}\right)^{-1}\right] \psi(t;\tau) \;,
\end{equation}
where we have introduced the representation
\begin{equation}\label{rhoRep}
 \hat\rho = \frac{\ui}{2\tilde\alpha} \frac{\partial}{\partial t} \;.
\end{equation}

For $\tau>b^2$, the evolution involves the Coulomb-type Hamiltonian with zero
energy (due to the boundary condition $\dot\rho(\infty)=1$) and we get the
solution
\begin{equation}\label{psit1}
 \psi(t;\tau) = \left(\frac{1-t}{1+t}\right)^{\ui\alpha} \frac1{1-t^2}
 N(\alpha,y) \;, \qquad (\tau > b^2) \;,
\end{equation}
where the normalization factor $N(\alpha,y)$ will be fixed later on. On the
other hand, for $\tau \leq b^2$ we have just free evolution,
\begin{equation}\label{freeEv}
 \ui\frac{\partial}{\partial \tau} \log\psi(t;\tau) = -\tilde\alpha(1-t^2) \;,
\end{equation}
yielding
\begin{equation}\label{psit2}
 \psi(t;\tau) = N(\alpha,y) \left(\frac{1-t}{1+t}\right)^{\ui\alpha}
 \frac1{1-t^2} \esp{\ui\alpha(1-\ui y)(1-t^2)(\tau-b^2)} \;, \qquad
 (\tau \leq b^2)
\end{equation}
and therefore
\begin{equation}\label{psiI}
 \psi(\rho;\tau) = N(\alpha,y) \int\dif t\;
 \left(\frac{1-t}{1+t}\right)^{\ui\alpha} \frac1{1-t^2}
 \esp{\ui\alpha(1-\ui y)(1-t^2)(\tau-b^2)}
 \esp{\ui\alpha(1-\ui y)\rho t} \;.
\end{equation}
Finally, by setting $\rho=0$ and $\tau=0$ we get the desired result, which
differs from the representation~(\ref{Trep}) by an integration by parts, by a
normalization factor and by the replacement $b^2\to b^2(1-\ui y)$.

We thus conclude that the elastic $S$-matrix (or, the tunneling amplitude
including absorption) is given by
\begin{equation}\label{SbsY}
 \S_\el(b,s,Y_b) = \frac{(\ui\alpha/\esp{})^{\ui\alpha}}{\Gamma(\ui\alpha)}
 \tanh(\pi\alpha) \,2 b^2(1- \ui y) N(\alpha,y) \int_{-1}^{+1} \dif t\; t\,
 \esp{\ui\alpha\F\big(t,b^2(1-\ui y)\big)} \;,
\end{equation}
where the factor $N$ is now computed by the ``Coulomb phase'' normalization
condition on $\S_\el(b=\infty)$ to be
\begin{equation}\label{NormFact}
 N(\alpha,y) = (1-\ui y)^{\ui\alpha} \;.
\end{equation}

\subsection{Perturbative versus collapse-like regimes\label{pvclr}}

\begin{figure}[!ht]
  \centering
  \includegraphics[width=0.45\textwidth]{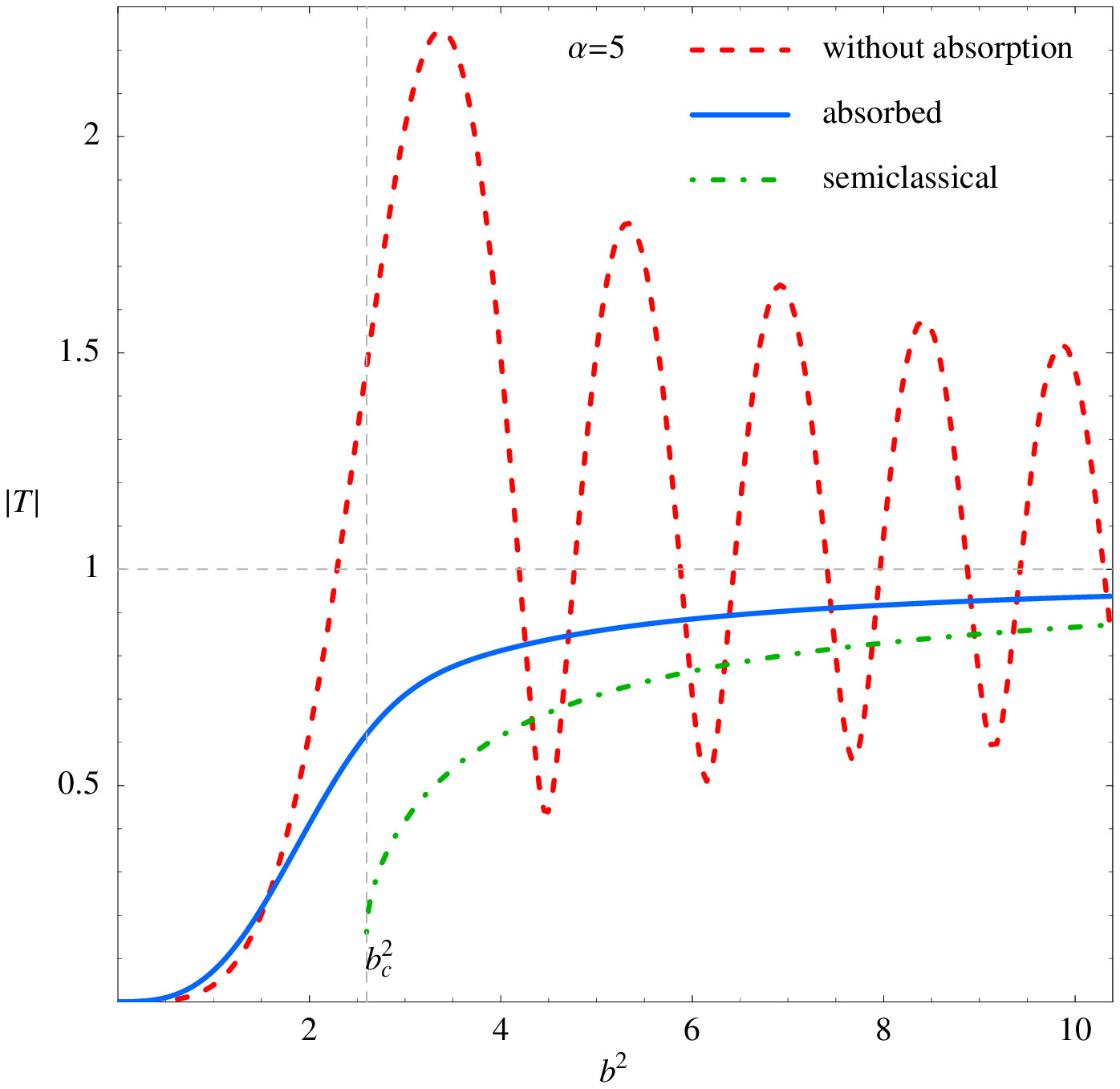}\hfill
  \includegraphics[width=0.45\textwidth]{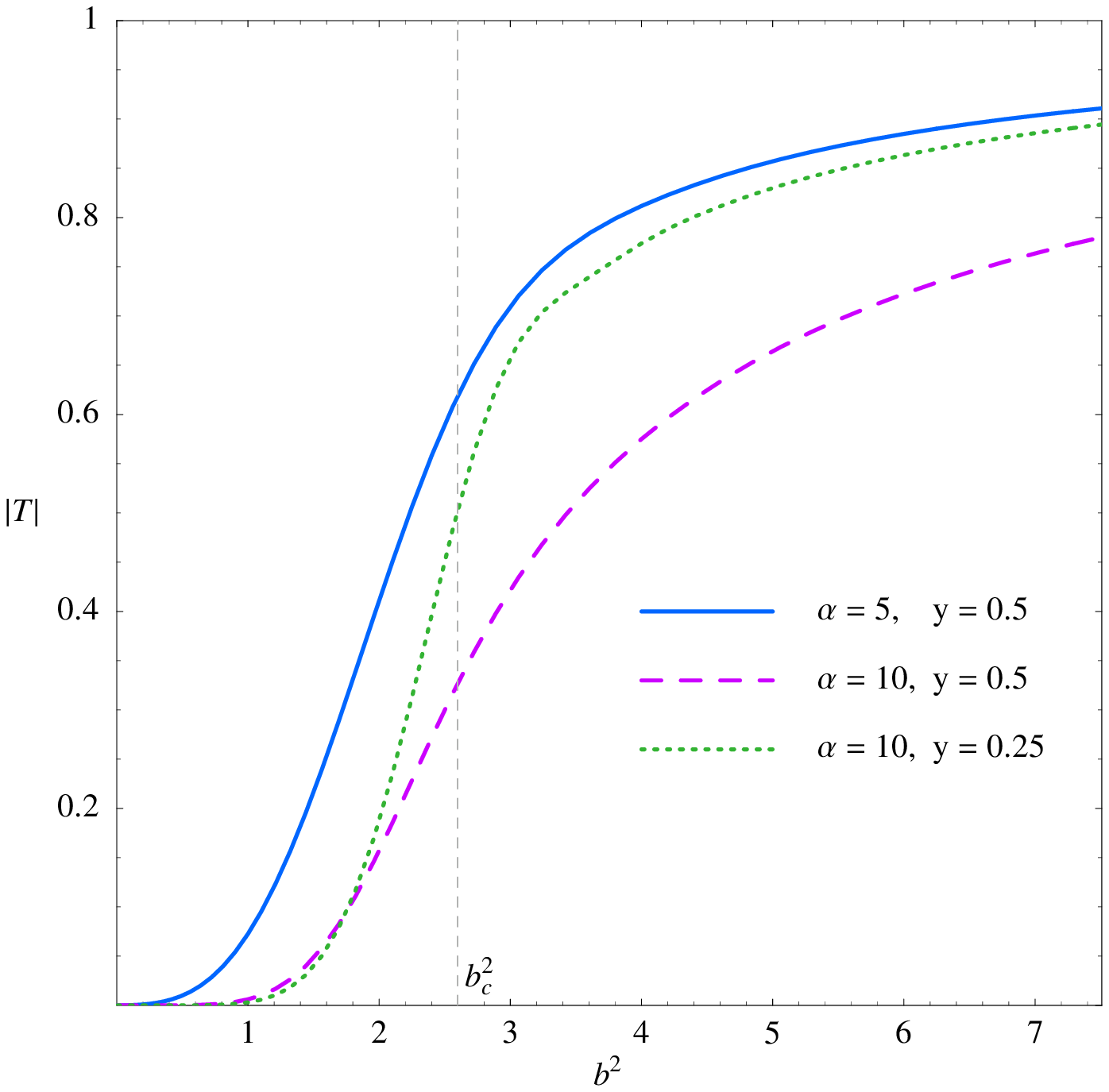}
  \caption{\it a) Tunneling amplitude with elastic absorption, for two values of
    the absorption parameter $y=0$ (dashed-red) and $y=0.5$ (solid-blue), the
    latter to be compared with the semiclassical result (dash-dotted green). b)
    Absorbed amplitudes for different values of $\alpha$ and $y$. (The two
    solid-blue curves in a) and b) are the same amplitudes, just on different
    scales).}
  \label{f:assorbElastico}
\end{figure}

The elastic $\S$-matrix resulting from eq.~(\ref{SbsY}) improves the
semiclassical approximation by providing a quantum weight to the various
classical paths. Its modulus is plotted in fig.~\ref{f:assorbElastico} for
various values of $\alpha$ and of the absorption parameter $y$.
Fig.~\ref{f:assorbElastico}a shows that the $y=0$ result oscillates, so that
absorption (required for self-consistency because $h\sim1-\dot\rho\neq 0$) is
essential to comply with elastic unitarity ($|\S_\el|\leq 1$).

The oscillations appearing in fig.~\ref{f:assorbElastico} in the $y=0$ limit are
due to the fact that the quantum treatment embodies contributions from both the
perturbative and the non-perturbative semiclassical solutions for $b>b_c$. In
the saddle point approximation one would have the simple formula
\begin{align}
 \S_\el(b,s) &= \sum_j \S_j(s) = \sum_j \esp{\ui\alpha\F(\tb_j,b^2)}
 \sqrt{\frac{2\tb_j}{3\tb_j^2-1}}{\cal J}(\tb_j,s)(1+\dots)
 \label{Selas} \\ \nonumber
 {\cal J}(\tb,s) &= \esp{-\frac{2Y_b \alpha}{\pi}\int\dif\tau\;
 (1-\dot\rho_\cl)^2}
 = \exp\left[-\alpha\frac{2Y_b}{\pi}\left(\frac1{\tb}-1\right)\right] \;.
\end{align}
If the absorption factor ${\cal J}(t_b,s)$ were neglected (as it happens in the
$y=0$ limit), there would be strong interference of the two solutions, and large
unitarity violations. For $y=0.5$, on the other hand, the non-perturbative
contribution is much more absorbed, and the result is dominated, for $b>b_c$, by
the perturbative contribution. This explains the overall agreement with the
semiclassical result --- also plotted in fig.~\ref{f:assorbElastico}a --- with
moderate quantum corrections. Finally, in fig.~\ref{f:assorbElastico}b we note
the dependence of absorption on $\alpha$ and $y$. While for $b\gg b_c$ it is
basically the product $y\alpha$ that matters, for $b<b_c$ absorption is $\alpha$
dependent, but only weakly $y$-dependent, showing a sort of universal behaviour
up to sizeable values of $y$.

The $S$-matrix behaviour for $b\sim b_c$ requires a special discussion.  When
$t_b$ in eq.~(\ref{Selas}) approaches the critical value $t_c=1/{\sqrt3}$ (or,
$b^2\simeq b^2_c=3\sqrt{3}/2$) at which the criticality equation is stationary,
the two dominant contributions to (\ref{Selas}) become of the same order, and
the saddle points pinch and then become complex conjugate (at $y=0$). In such a
situation the quadratic fluctuations diverge, and the quadratic expansion of the
phase~(\ref{phase}) has to be extended to the cubic terms, in order to stabilize
the integration.

If absorption is included, this analysis has to be performed at complex values
of $\tilde{b}^2\equiv b^2(1-\ui y)$, by assuming that
$\tilde{\beta}\equiv\frac{\tilde{b}^2}{b_c^2}-1\equiv \beta -\ui y$ is a small
parameter (see app~\ref{a:spe}). We expand the phase in (\ref{Trep}) around the
value $t_m(\tilde{b}) $ such that $\F''(t_m,\tilde{b}^2)=0$, which is given
precisely by eq.~(\ref{chimin}) and acquires the meaning of WKB turning point.
To lowest order in $\tilde{\beta}$ we find the expression
\begin{equation}\label{cubicexp}
 \F(t, \tilde{b}^2)=\F_c -\sqrt 3 \tilde{\beta}+3\tilde{\beta}(t-t_m)
 -\frac92(t-t_m)^3+\cdots
\end{equation}
which, replaced in eq.~(\ref{SbsY}), yields the integral representation of an
Airy function (cf.\ eq.~(\ref{airyRep})), by providing the approximate result
\begin{align}\label{Airy}
 \S_\el &\sim \esp{-\alpha(\ui \sqrt 3\tilde{\beta}-y)}
 \Ai(-2^{1/3}\alpha^{2/3}\tilde{\beta}) \simeq
 \esp{-\alpha\left(\ui\sqrt 3\tilde{\beta}-y +\frac{2\sqrt2}3
 (-\tilde{\beta})^{3/2}\right)}
\end{align}
We see that the large-$\alpha$ behaviour is characterized by the
$(-\beta+\ui y)^{3/2}$ exponent for $y>0$ also. However, the modulus $|\S_\el|$
behaves differently for positive and negative $\beta$'s. Indeed, a simple
evaluation yields
\begin{align}
 |\S_\el| &\simeq \esp{-\alpha y(\sqrt{3}-1)} \,
 \esp{\sqrt{2}\alpha y\sqrt{\beta}} & (\beta&\gg y) \nonumber \\
 &\simeq \esp{-\alpha y(\sqrt 3-1)} \,
 \esp{-\frac{2\sqrt2}{3}\alpha(-\beta)^{3/2}} & (-\beta&\gg y)
\end{align}
Therefore, the additional absorption with exponent $3/2$ for
$b<b_c\;\;(\beta<0)$ is unambiguously confirmed, and is continuously joined to
the $y\sqrt{\beta}$ behaviour of normal absorption for $b>b_c$. This provides a
rise of the modulus in the small $\beta>0$ region, so as to match eventually the
(small) perturbative absorption at large values of $b$. Such features
qualitatively explain the behaviour of absorption in the $b\sim b_c$ region, as
pictured in fig.~\ref{f:assorbElastico}b, in particular the fact that, for
$b<b_c$, it is weakly $y$-dependent.

The above analysis confirms the existence of a perturbative and a collapse-like
regime, and also confirms the singularity of exponent $3/2$ for the asymptotic
high-energy behaviour. However, we also see from eq.~(\ref{Airy}) that
$\tilde{b}=b_c$ {\it is not} a real singularity of the $S$-matrix, because the
Airy function is an entire function of its argument $\sim \tilde{\beta}$. This
means that the singularities due to the pinching solutions cancel each other at
$b=b_c$. The lack of a true singularity might favour the interpretation of the
new regime as a collective phenomenon, rather than a signal of new states.
However, since our quantization procedure is incomplete, it is not yet clear
whether such a feature is kept at full quantum level, in the string-gravity
framework.

\section{Discussion}

Here we have introduced a quantization procedure for the transverse-space
dynamics of the ACV framework~\cite{ACV07}, which allows a deeper understanding
of the transplanckian scattering matrix of light particles and of its
high-energy regimes, by providing a quantum tunneling interpretation of the
collapse-like regime $b\lesssim R=4G\sqrt s$.

Indeed, we have related the high-energy elastic $S$-matrix to a tunneling
amplitude in gravitational field space ($h(r^2)\equiv \nabla^2\phi$) from a
weak-field configuration at large distances ($h(\infty)=0$) to a regular field
at short distances ($h(0)$ finite). The dynamics, embodied in the high-energy
graviton emission vertex, provides an impact-parameter dependent Coulomb barrier
in the ``renormalized radius'' $\rho\equiv r^2(1-(2\pi R)^2\dot{\phi})$. As a
consequence, if the impact parameter $b$ is below some critical value $b_c\sim
R$, quantum tunneling is essential for the above configuration to occur and
provides a calculable suppression of the elastic channel which corresponds to
the complex semiclassical solutions of the ACV proposal~\cite{ACV07}.

The statement above summarizes the basic point that we meant to elucidate here.
Furthermore, our quantum approach provides a deeper perspective on a number of
other points.
\begin{itemize}
\item Introducing ``normal'' absorption (due to graviton emission) is essential
  for the $S$-matrix to satisfy elastic unitarity, and in order to understand
  the relative weight of the various semiclassical solutions and, in particular,
  the dominance of the perturbative one for $b\gg b_c$.
\item Taking absorption into account amounts to consider a complex impact
  parameter $\tilde{b}^2\equiv b^2(1-\ui y)$, where $y=2Y_b/\pi$ parametrizes
  the rapidity phase space allowed by energy conservation~\cite{CV08} for
  graviton emission. This makes it possible to continue normal absorption below
  $b=b_c$, thus matching the perturbative behaviour at large $b$ values to the
  ``extra'' tunneling suppression for $b<b_c$. The latter is thus fully
  confirmed, and shows a sort of universality, in the sense that it is weakly
  $y$-dependent.
\item Quantum corrections are calculable, of relative order $1/\alpha$ and tend
  to smooth out the semiclassical results. In particular, $b=b_c$ is a
  singularity of the asymptotic high-energy behaviour, but {\it is not} a
  singularity of the $S$-matrix itself.
\end{itemize}

In the present paper we investigate the elastic $S$-matrix only. Although
absorptive effects are consistently taken into account and are essential to
fulfill elastic unitarity, we do not consider inelastic matrix elements
explicitly. Our quantum model can be extended, in principle, to inelastic
channels according to eq.~(\ref{inSpi}), by introducing some time-dependent
external current which, however, makes the inelastic model no longer explicitly
solvable. Furthermore, improvements are needed, for instance in connection with
energy conservation~\cite{CV08}. For the above reasons a deeper analysis is
desirable, and is deferred to further work. Hopefully, the outcome of such a
work should be able to explain the mechanism by which inelastic unitarity can be
satisfied.

\section*{Acknowledgements}

We warmly thank Daniele Amati and Gabriele Veneziano for seminal discussions on
the tunneling interpretation and Slava Rychkov for stimulating remarks on the
role of multigraviton emission. Work supported in part by a PRIN grant (MIUR,
Italy).

\appendix

\section{Stationary-phase estimates of tunneling amplitude\label{a:spe}}

In this appendix we provide analytic approximations for the tunneling
amplitude~(\ref{Trep}) and the elastic scattering amplitude~(\ref{SbsY}) when
$\alpha$ is a large parameter. This amounts to evaluate the $t$-integral
\begin{equation}\label{tint}
 I(\alpha,b^2) \equiv \int_{-1}^{+1} \dif t\; t\,
 \esp{\ui\alpha\F(t,b^2)}
\end{equation}
in an approximate form, with $\F$ given by eq.~(\ref{phase}).

\subsection{Quadratic expansion}

Let us first look for stationary points $\tb$ of the phase $\cal F$ given in
eq.~(\ref{phase}):
\begin{equation}\label{Fp0}
 \F' \equiv \partial_t\F(t,b^2) = 2\left(b^2 t-\frac1{1-t^2}\right)
 = 0 \qquad \iff \qquad \tb(1-\tb^2) = \frac1{b^2}
\end{equation}
The stationary points are thus determined by the criticality
condition~(\ref{critical}), and for $b>b_c$ the
integration interval $t\in[-1,1]$ contains two of the three real solutions:
\begin{align}
 \tb_1 &= 1-\frac1{2b^2}+\ord{\frac1{b^4}} &&(\text{perturbative solution}) \\
 \tb_2 &= 0+\frac1{b^2}+\ord{\frac1{b^4}}  &&
(\text{non-perturbative solution}) \;.
\end{align}
The second derivative of the phase at the stationary points is given by
\begin{equation}\label{Fss}
 \F'' \equiv \partial^2_t\F(t,b^2) =
 2\left(b^2-\frac{2t}{(1-t^2)^2}\right) \;, \qquad
 \F''(\tb,b^2) = -2 b^4 \tb (3\tb^2-1)
\end{equation}
and increases with $b$:
$\F''(\tb_1)\sim b^4$, $\F''(\tb_2)\sim b^2$. Therefore, the
fluctuations around the two saddle points are smaller and smaller at large $b$,
while the distance between them increases, allowing us to treat them separately
for sufficiently large $b$:
\begin{align}
 I(\alpha,b^2) &\simeq \int_{-1}^{1}\dif t\; t \sum_{j=1}^2
 \esp{\ui\alpha[\F(\tb_j)+\frac12\F''(\tb_j)(t-\tb_j)^2]}
 \simeq \sum_{j=1}^2 \tb_j \esp{\ui\alpha\F(\tb_j)}
 \sqrt{\frac{2\pi\ui}{\alpha\F''(\tb_j)}} \\
 &\simeq \frac1{b^2} \sqrt{\frac{\pi}{\ui\alpha}}
 \sum_{j=1}^2 \esp{\ui\alpha\F(\tb_j)} \sqrt{\frac{\tb_j}{3\tb_j^2-1}} \;.
\end{align}
The saddle-point estimate of the tunneling amplitude is therefore given by
\begin{align}
 \T(b,\alpha) &\simeq
 \frac{(\ui\alpha/\esp{})^{\ui\alpha}}{\Gamma(\ui\alpha)}
 \tanh(\pi\alpha) \,2 \sqrt{\frac{\pi}{\ui\alpha}}
 \sum_{j=1}^2 \esp{\ui\alpha\F(\tb_j)} \sqrt{\frac{\tb_j}{3\tb_j^2-1}}
 \nonumber \\
 &\simeq \sum_{j=1}^2 \esp{\ui\alpha\F(\tb_j)}
 \sqrt{\frac{2\tb_j}{3\tb_j^2-1}} \;,
 \label{Tsp}
\end{align}
where in the last equality we have used the Stirling approximation for
$\Gamma(\ui\alpha)$ valid when $\alpha\gg 1$. The result~(\ref{Tsp}) reproduces
the elastic $S$-matrix result eq.~(\ref{Selas}) without absorption
(${\cal J}=1$).

\subsection{Cubic expansion}

When the impact parameter $b$ approaches the critical value $b_c$, the two
gaussians centered at the saddle points overlap and cannot be considered
separately. In particular, at $b=b_c$ the two saddle points coincide
\begin{equation}\label{tc}
 \tb_1(b_c^2) = \tb_2(b_c^2) = t_c = \frac1{\sqrt{3}}
\end{equation}
and have infinite fluctuations, due to the vanishing of the second derivative
$\F''(t_c,b_c^2)=0$. Therefore, an estimate of the integral~(\ref{tint}) for
$b\simeq b_c$ requires an expansion of the phase $\F$ up to the third order in
$t$. In order to exploit the Airy-function integral representation
\begin{equation}\label{airyRep}
 \Ai(z) = \int_{\esp{-\ui\pi/3}\infty}^{\esp{+\ui\pi/3}\infty}
\frac{\dif\xi}{2\pi\ui} \; \esp{-z\xi+\frac{\xi^3}{3}} \;,
\end{equation}
we expand $\F$ around the point $t_m(b^2)$ where its second derivative vanishes:
\begin{equation}\label{Fs0}
 \F''(t_m,b^2) = 0 \qquad\iff\qquad \frac{t_m}{(1-t_m^2)^2} = \frac{b^2}{2} \;,
\end{equation}
while the third derivatives is given by
\begin{equation}\label{Fterzo}
 \F''' \equiv \partial^3_t\F(t,b^2) = -4\frac{1+3t^2}{(1-t^2)^3} \;.
\end{equation}
By denoting with the subscript $m$ the various quantities evaluated at $t=t_m$,
and by setting $\xi\equiv\ui[\alpha(-\F'''_m)/2]^{1/3}(t-t_m)$ we obtain
\begin{align}
 I(\alpha,b^2) &\simeq \int\dif t\;t \exp\left\{\ui\alpha\left[
 \F_m + \F'_m (t-t_m) + \frac16 \F'''_m (t-t_m)^3 \right]\right\}
 \nonumber \\
 &\simeq t_m \esp{\ui\alpha\F_m} 2\pi
 \left(\frac{2}{-\F'''_m\alpha}\right)^{1/3}
 \Ai\left(\alpha^{2/3}\F'_m\Big(\frac{2}{-\F'''_m}\Big)^{1/3}\right) \;.
\end{align}
In order to express the above results in terms of the impact parameter $b$, it
is convenient to introduce the (small) parameter $\beta$ such that
\begin{align}
  b^2 = b_c^2 (1+\beta) \;.
\end{align}
From eqs.~(\ref{Fs0},\ref{phase},\ref{Fp0},\ref{Fterzo}) we have
\begin{align}
  t_m &= t_c\left(1+\frac{\beta}{3}+\ord{\beta^2}\right) \;, \qquad
  \tb{} = t_c\left( 1 \pm \sqrt{\frac{2}{3}\beta} + \ord{\beta^2} \right)
  \nonumber \\
 \F_m &= \F_c - \sqrt{3}\beta + \ord{\beta^2} \;, \qquad
 \F_c \equiv \F(t_c,b_c^2) = \log(4L^2)-\log(2+\sqrt{3})+1-\sqrt{3} \nonumber \\
 \F_m' &= 3\beta + \frac{\beta^2}{2} + \ord{\beta^3}
 \nonumber \\
 \F_m''' &= -27 \left(1 + \frac{4\beta}{3} + \ord{\beta^2}\right)
\end{align}
and finally
\begin{equation}\label{Tairy}
 \T(\alpha,b^2) \simeq 2^{11/6} \, 3^{-3/2} \, \sqrt{\ui\pi} \, \alpha^{1/6} \,
 b^2 \, \esp{\ui\alpha\F_c} \, \esp{-\ui\alpha\sqrt{3}\beta} \,
 \Ai(-2^{1/3}\alpha^{2/3}\beta) \;.
\end{equation}

The behaviour of the elastic scattering amplitude~(\ref{SbsY})
\begin{equation}\label{SelT}
 \S_\el(b^2,s,Y_b) = (1-\ui y)^{\ui\alpha} \T\big(b^2(1-\ui y),\alpha\big)
\end{equation}
for $b\simeq b_c$ and small $y$ can be obtained directly from eq.~(\ref{Tairy}).
In fact, if we denote
\begin{equation}\label{btilde}
 \tilde{b}^2 \equiv b^2(1-\ui y) \equiv b_c^2( 1+\tilde\beta ) \;,
\end{equation}
the complex parameter
\begin{equation}\label{betatilde}
 \tilde\beta = \beta - \ui y(1+\beta) \simeq \beta - \ui y
\end{equation}
is small. The factor $\T\big(b^2(1-\ui y),\alpha\big)$ is then given by the
expression~(\ref{Tairy}) with $\beta\to\tilde\beta$. The remaining factor can be
rewritten as
\begin{equation}\label{Nfact}
 N(\alpha,y) = (1-\ui y)^{\ui\alpha} \simeq (\esp{-\ui y})^{\ui\alpha}
 = \esp{\alpha y} \qquad ( y \ll 1 )
\end{equation}
and we obtain
\begin{equation}\label{Sairy}
 \S_\el(b^2,s,Y_b) \simeq 2^{11/6} \, 3^{-3/2} \, \sqrt{\ui\pi} \,
 \alpha^{1/6} \, \tilde{b}^2 \, \esp{\ui\alpha\F_c} \,
 \esp{-\alpha(\ui\sqrt{3}\tilde\beta-y)} \,
 \Ai(-2^{1/3}\alpha^{2/3}\tilde\beta) \;,
\end{equation}
from which we derive eq.~(\ref{Airy}). The r.h.s.~of eq.~(\ref{Airy}) is
obtained from the asymptotic behaviour
\begin{equation}\label{AiryAsy}
 \Ai(z) \simeq \frac1{2\sqrt{\pi}z^{1/4}}\exp\left(-\frac{2}{3}z^{3/2}\right)
 \qquad (z\to\infty \;, \quad |\arg(z)| < \pi) \;,
\end{equation}
where we can consider the argument of $\Ai$ to be large even for small
$\tilde\beta$, provided $\alpha$ is large enough:
\begin{equation}\label{zlarge}
 |z| = 2^{1/3} \alpha^{2/3}\,|\tilde\beta| \gg 1 \;.
\end{equation}


\end{document}